\documentclass[onecolumn,showpacs,preprintnumbers,showkeys,superscriptaddress]{revtex4}
\usepackage{hyperref}
\usepackage{amssymb}
\usepackage{amsmath}

\usepackage{color}
\usepackage{tipa}
\usepackage{graphicx}
\usepackage{dcolumn}
\usepackage{bm}

\def\eqref#1{Eq.~(\ref{eq:#1})}

\begin{document}

\title{Lowest eigenvalue of the nuclear shell model Hamiltonian}
\author{J. J. Shen}
\affiliation{Department of Physics,  Shanghai Jiao Tong University,
Shanghai 200240, China} \affiliation{Nishina Center, RIKEN, the
Institute of Physical and Chemical Research, Hirosawa 2-1, Wako-shi,
Saitama 351-0198, Japan }
\author{Y. M. Zhao}     \email{ymzhao@sjtu.edu.cn}
\affiliation{Department of Physics,  Shanghai Jiao Tong University,
Shanghai 200240, China}  \affiliation{Center of Theoretical Nuclear
Physics, National Laboratory of Heavy Ion Accelerator, Lanzhou
730000, China} \affiliation{CCAST, World Laboratory, P.O. Box 8730,
Beijing 100080, China}
\author{A. Arima}
\affiliation{Department of Physics, Shanghai Jiao Tong University,
Shanghai 200240, China}\affiliation{Science Museum, Japan Science
Foundation, 2-1 Kitanomaru-koen, Chiyoda-ku, Tokyo 102-0091, Japan}

\date{\today}

\begin{abstract}
In this paper we investigate regular patterns of matrix elements of
the nuclear shell model Hamiltonian $H$, by sorting the diagonal
matrix elements from the smaller to larger values.  By using simple
plots of non-zero matrix elements and lowest eigenvalues of
artificially constructed ``sub-matrices" $h$ of $H$, we propose a
new and simple formula which predicts the lowest eigenvalue with
remarkable precisions.
\end{abstract}


\vspace{0.4in}

\maketitle

\newpage

The diagonalization of  matrices is a fundamental practice in
nuclear structure physics as well as many other fields. However,
diagonalization becomes difficult if the dimension of the matrix is
very large. Statistical approaches are very suggestive and have been
developed, e.g., in Refs. \cite{K.F.Ratcliff, F.J.Margetan, V,
Papenbrock1, Papenbrock2, Yoshinaga1, Shen1}, where the lowest
eigenvalue is presented in terms of  the energy centroid and
spectral moments.

Recently, we showed in Ref. \cite{Shen2} that sorting  diagonal
matrix elements of a given nuclear shell model Hamiltonian from the
smaller to the larger values provides us with a new approach to
evaluate the eigenvalues. By sorting the diagonal matrix elements,
we are able to evaluate all eigenvalues based on a very strong
linear correlation between the diagonal matrix elements and exact
eigenvalues. This method was found to work very well for medium
eigenvalues but deviates for the lowest ones. However, in nuclear
structure physics as well as many other fields of sciences, we are
interested in the low-lying states. It is therefore very desirable
to refine the approach towards more and more accurate evaluation of
the low-lying eigenvalues, by sorting the diagonal matrix elements.

In this paper, we propose a new approach to predict the lowest
eigenvalue of the nuclear shell model Hamiltonian. To exemplify our
method, we shall use a few realistic examples of nuclear shell
model calculations. All results in this article are based on the
shell model code by the Kyushu group \cite{Takada,Takada2,Takada4}. The
shell model basis states of this code are constructed by using the
coefficients of fractional parentage discussed in Ref.
\cite{Takada2}. In this paper we take the USD interactions of Ref.
\cite{USD}. Other interactions such as the Yukawa-type interactions
of Refs. \cite{Arima,Takada4} give similar results.

\begin{figure}
\includegraphics[width= 0.65\textwidth]{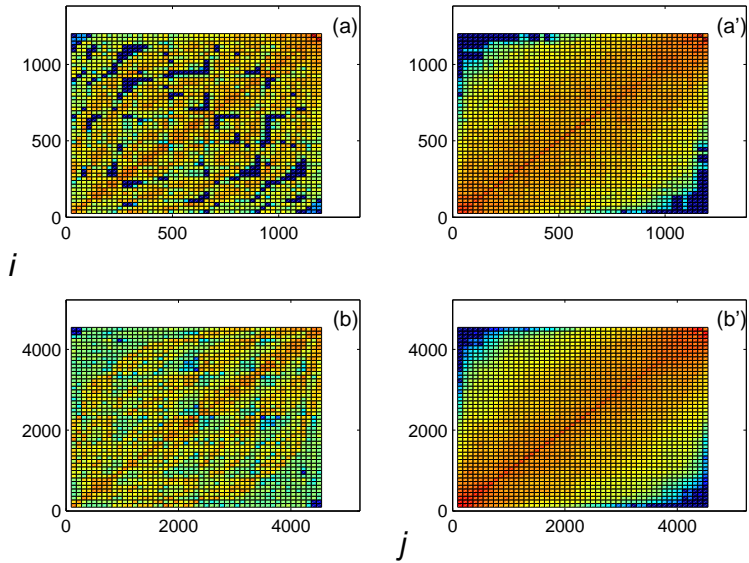}
\caption{(Color online)  Magnitude of the matrix elements without
(panels (a) and (b)) and with (panels (a$'$) and (b$'$)) sorting the
diagonal matrix elements from the smaller to the larger. The color
from blue to red corresponds to values from zero to large
magnitudes. The results are based on $I^{\pi}=0^+$ (panels a and
a$'$) and $I^{\pi}=2^+$ (panels b and b$'$) states of the $^{24}$Mg
nucleus, obtained by using the USD interactions. The magnitude of
$H_{ij}^{(I)}$ without sorting the diagonal matrix elements (left
hand side) are close to ``random", and those with sorting the
diagonal matrix elements (right hand side) decrease rapidly as going
farther from the diagonal line. }\label{geometry} \end{figure}

Let us denote the matrix of spin $I$ states of the nuclear shell
model Hamiltonian $H$ by $H^{(I)}$, and the matrix elements of
$H^{(I)}$ by  $H_{ij}^{(I)}$, where $i$ and  $j$ represent indices
of basis states. In Fig. \ref{geometry}, we present two typical
examples of distributions of the magnitude of $H_{ij}^{(I)}$, based
on the $J^{\pi}=0^{+}$ and $J^{\pi}=2^{+}$ states of the $^{24}$Mg
nucleus. The color from blue to red corresponds to values from zero
to large magnitudes. From panels (a-b) of Fig. \ref{geometry}, one
sees that the values of $H_{ij}^{(I)}$ (panels (a-b) of Fig.
\ref{geometry}) look ``random". However, if one sorts the diagonal 
matrix elements from the smaller to larger values, as in Refs.
\cite{Shen2}, the values of $H_{ij}^{(I)}$ decrease rapidly and
become zero if they are ``far" enough from the diagonal line,  as
shown in panels (a$'$-b$'$) of Fig. \ref{geometry}.

\begin{figure}
\includegraphics[angle=0,width=0.75\textwidth]{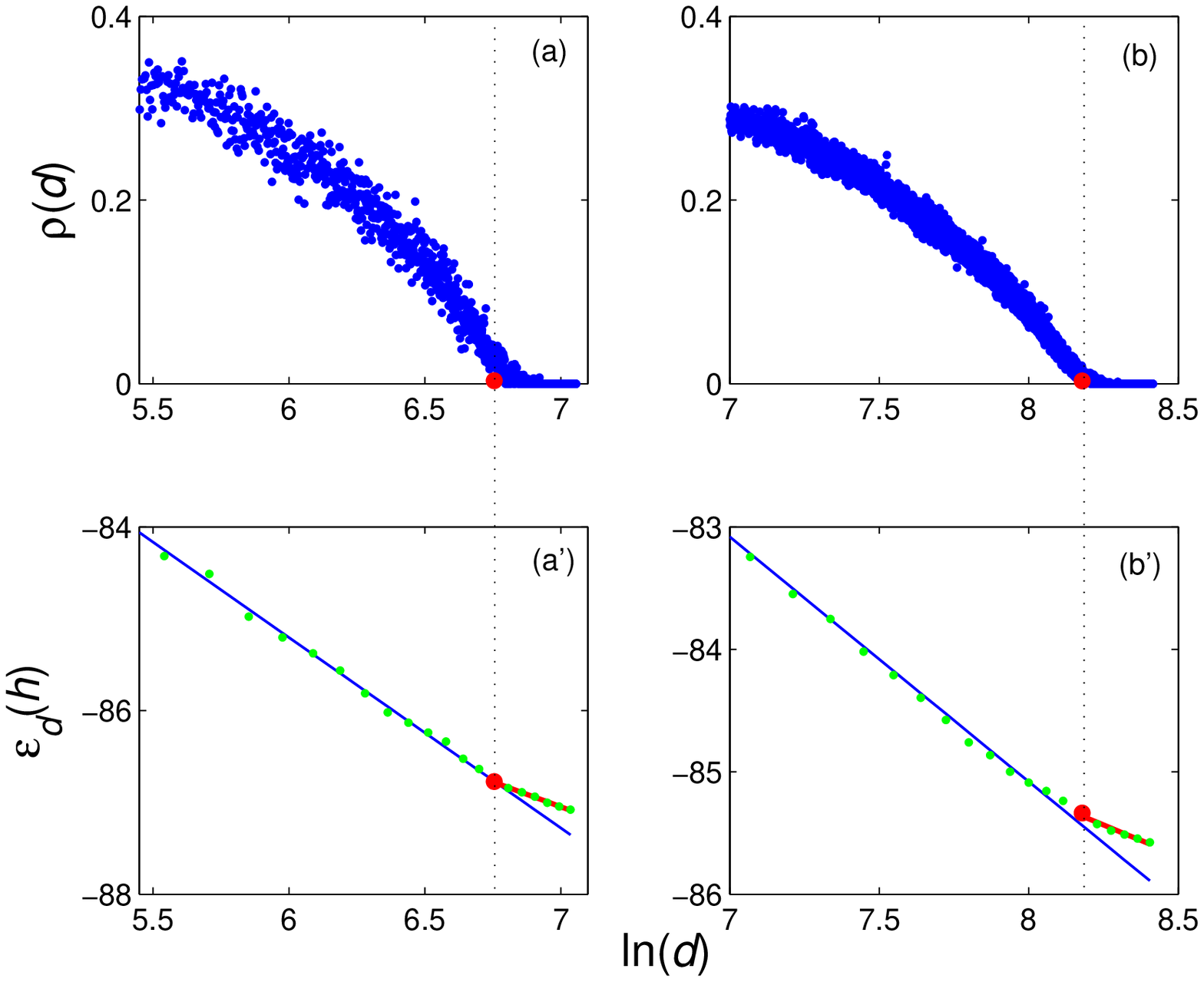}
\caption{~ (Color online) ~ Distribution for $H_{ij}^{(I)}$ to be
non-zero, denoted by $\rho(d)$, and the lowest eigenvalues of $h$,
denoted by $\epsilon_d(h)$. The results of (a,a$'$) and (b,b$'$) are
obtained  for the $J^{\pi}=0^{+}$ and $J^{\pi}=2^{+}$ states of
$^{24}$Mg, respectively. One sees that the values of $d$ where $\rho(d_0)=0$
and where the $\epsilon_d(h)$-ln$\,d$ plots change the slope ($D_0$)
coincide, i.e., $d_0=D_0$. We use the same scale in (a,a$'$) or
(b,b$'$). The dotted lines are used to guide the eyes. }\label{point}
\end{figure}

Let us investigate this behavior in another form. We study the
probability for $H_{ij}^{(I)}$ to be non-zero (after sorting the
diagonal matrix elements of $H_{ij}^{(I)}$), while moving away from the
diagonal line, versus $d$, denoted by $\rho (d) = \frac{\sum{|{\rm
sgn}(H_{i,i+d}^{(I)})|}}{D-d},$ $d=1, 2, 3, \cdots, D$. Here $d$ is the
``distance" of $H_{ij}^{(I)}$ from the diagonal line, and $D$ is the
dimension of matrix $H_{ij}^{(I)}$ for spin $I$ states.  As shown in
Fig. \ref{point}(a) and (b) for the $I^{\pi}=0^+$ and $2^+$ states
of the $^{24}$Mg nucleus,  $\rho(d)$ becomes zero at a critical
value $d=d_0$;  the value  of ln$\,d_0$ equals 6.75 and 8.22,
respectively.

An argument for the regular patterns described in  Fig. 1 and Fig.2
(a) and (b) is as follows. With the diagonal matrix elements sorted
from the smaller to larger values, one classifies configurations
from the lowest to the largest in energy, roughly by particle-hole
excitations. Configurations that come first are  the lowest, and the
states that come last are $n$-particle-$n$-hole excitations out of
those low configurations. Because the shell-model Hamiltonian
consists of one-body and two-body operators, one can not connect
those configurations that are ``distant", e.g., the configurations
with the lowest energy and $n$-particle-$n$-hole configurations with
$n>2$. This explains the reason why all values of $H_{ij}^{(I)}$
become zero for $d \ge d_0$. Soon we shall find that the value of
$d_0$ is very important in predicting the lowest eigenvalue of the
matrix $H^{(I)}$.

In Refs. \cite{V,Yoshinaga1,Shen1} the lowest eigenvalue is
presented in terms of ln$D$, where $D$ is the dimension of the
matrix $H^{(I)}$. Although the formulas of the lowest eigenvalues
presented in Refs. \cite{V,Yoshinaga1,Shen1} are applicable to the
random ensemble average (not to {\it individual} sets of interactions
parameters), one naturally asks whether or not certain plots of the
lowest eigenvalue versus the dimension could be useful in evaluating
the lowest eigenvalue of realistic systems studied in this paper.
Let us  sort the diagonal matrix elements from the smaller to the
larger, as in Refs. \cite{Shen2}. Then we truncate artificially the
matrix $H^{(I)}$ and obtain a ``sub-matrix" $h$ with dimension $d$
($d< D$), and $h_{ij}  = H_{ij}^{(I)}$ ($i,j = 1, 2, \cdots, d$). We
diagonalize $h$ and  obtain the lowest eigenvalue $\epsilon_d$ of
the matrix $h$, and plot $\epsilon_d$ versus ln$\,d$. In Fig.
2(a$'$-b$'$) we present the $\epsilon_d$-ln$\,d$ plots for the
$I^{\pi}=0^+$ and $2^+$ states of the $^{24}$Mg nucleus. One sees
that $\epsilon_d$ decreases linearly with ln$\,d$ when $d$ is smaller
than a critical dimension $d=D_0$, and decrease again linearly with
ln$\,d$ but with a smaller slope. Apparently, the value of $D_0$ and
the slopes for both $d<D_0$ and $d>D_0$ suffice for the evaluation of
the lowest eigenvalue of $H_{ij}^{(I)}$.

In Fig. 2 one sees that the values of $d_0$ where $\rho (d_0) =0$
in Fig. 2(a) and (b) coincide with $D_0$ in Fig. 2(a$'$) and (b$'$),
respectively. Panels (a,b) are based on the same matrices as
(a$'$,b$'$), respectively. For convenience, we use the same scale in
panels (a,a$'$)  and (b,b$'$), and plot two dotted lines to guide the
eyes in order to see such coincidence. From Fig. 2 one also sees
that the slope for $d>d_0$ (denoted by $k'$) is smaller than that
for $d<d_0$ (denoted by $k$).

An intuitive understanding of the facts that  $d_0 \simeq D_0$ and
$k'<k$ is given as follows. Because $H_{ij}^{(I)}$ are zero when $d
> d_0$ (i.e., $\rho(d_0) = 0$ for $i,j>d_0$), there is no contribution to
the lowest eigenvalue from these matrix elements. On the other hand,
some of matrix elements $H_{ij}^{(I)}$ and $H_{ji}^{(I)}$, with $0<
i < D-d_0$ and $j> D-d_0$, are non-zero (see Fig. 1(a$'$) and
(b$'$)), lowering down the smallest eigenvalue of the $h$ matrix. The
slope of the $\epsilon_d$-ln$\,d$ plot therefore changes at $d=d_0$ and
becomes smaller for $d>d_0$ than that of $h$ for $d<d_0$.

\begin{figure}
\includegraphics[angle=0,width=0.75\textwidth]{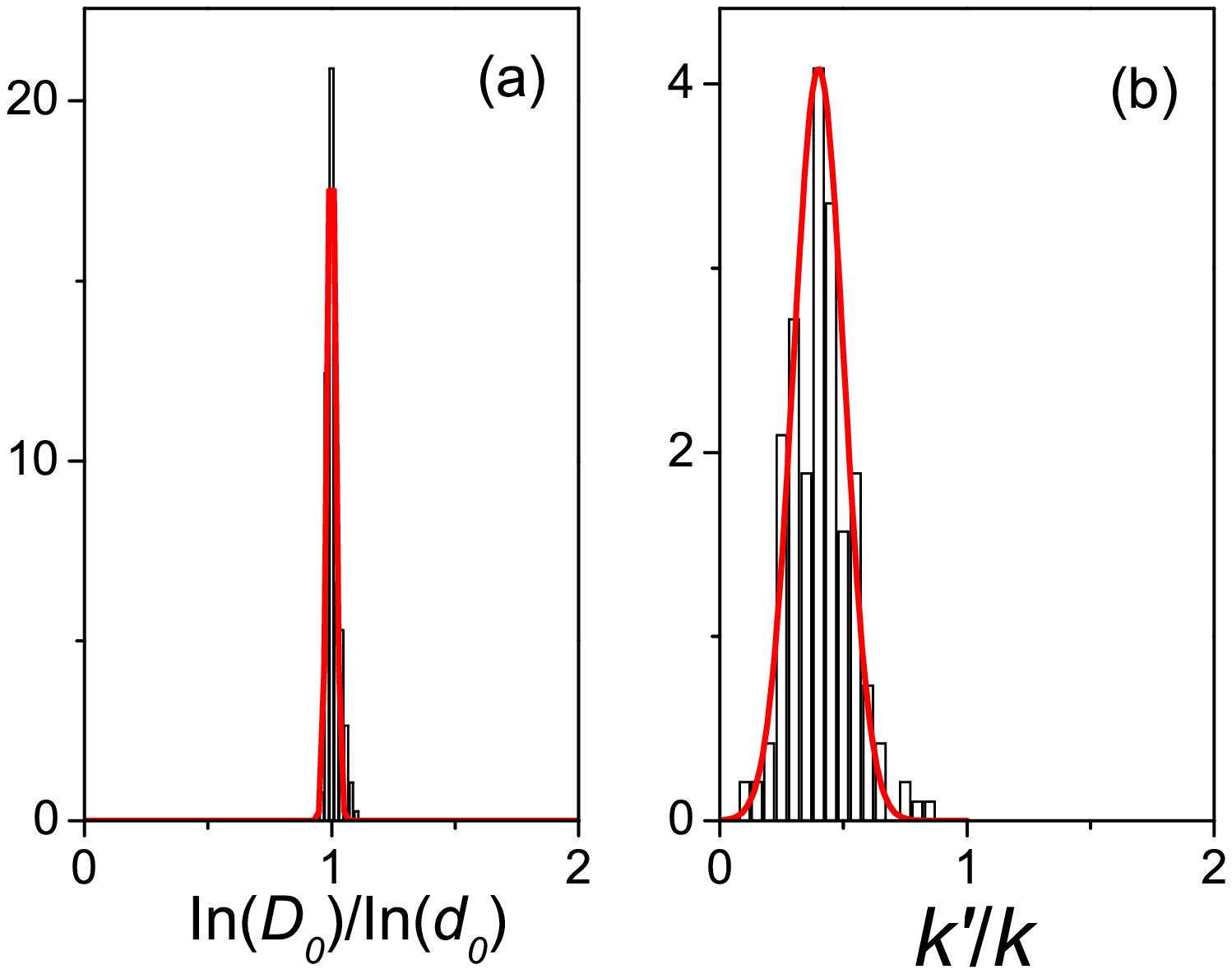}
\caption{~ (Color online) ~ Distribution for ln($D_0$)/ln($d_0$) and
$k'/k$ for two hundred  $H^{(I)}$ matrices for a number of nuclei in
the sd shell, the dimension $D$ of which goes from 500 to 10000, by
taking the USD interactions. One sees that ln($D_0$)/ln($d_0)\simeq
1$, and that $k'\sim \frac{2}{5}k $ with considerable fluctuations.
In this paper we assume that ln($D_0$)/ln($d_0$)=1 and
$k'=\frac{2}{5}k$ when we predict the lowest eigenvalue of
$H^{(I)}$, for simplicity. }
\end{figure}

In Fig. 3 we present our results of ln($D_0$)/ln($d_0$) and $k'/k$,
based on 200 examples of $H^{(I)}$ for a number of $sd$ shell
nuclei by using the USD interactions. One  sees that
ln($D_0$)/ln($d_0$) are very close to 1.0  and that $k'\simeq
\frac{2}{5}k$ with fluctuations. For simplicity we assume that
ln($D_0$)/ln($d_0$)=1 and $k'=\frac{2}{5}k$ for {\it all cases}
throughout this paper.

Making use of these regularities, we obtain a new and simple formula
to evaluate the lowest eigenvalue of the matrix $H^{(I)}$:
\begin{eqnarray}
E_{\rm min}^{(I)} & =&  \frac{2k}{5} {\rm ln} D + \left( k {\rm ln}
d_0 + b  -
\frac{2k}{5} {\rm ln} d_0 \right) \nonumber \\
& = & \frac{2k}{5} {\rm ln} (D)+ \frac{3k}{5} {\rm ln} (d_0)   + b ,
 \label{linear}
\end{eqnarray}
where $k$ and $b$ are the slope and intercept of the
$\epsilon_d$-ln$\,d$ plot for $d<d_0$; $d_0$ is determined by
$\rho(d_0)=0$, and $D$ is the number of spin $I$ states. Because the
$\epsilon_d$-ln$\,d$ plot shows a nice linearity (see Fig.
2(a$'$,b$'$)), we extract the values of $k$ and $b$ based on
sub-matrices $h$ of $H^{(I)}$ with $d \le D/10$ for all cases.

In Fig. 4 we present a comparison of the lowest states of spin $I$
predicted by the above Eq. (\ref{linear}), and those obtained by
the linear correlation (i.e., Eq. (3) of Ref. \cite{Shen2}), with
those calculated by diagonalizing $H^{(I)}$ ($I=0, 2, 4, 6, 8$) for
two nuclei, $^{24}$Mg and $^{28}$Si. One sees the remarkable
agreement between the exact eigenvalues (the column ``exact") and
our predicted ones (``pred1"), and substantial improvements achieved
by Eq. (\ref{linear}) in comparison with Eq. (3) of Ref.
\cite{Shen2} (``pred2" in Fig. 4). Without going into details, we mention
that the
overall root-mean-squared deviation ${\cal E}$ for the two-hundred
cases we checked in Fig. 3 (defined by ${\cal E}=  \sum_i^N \sqrt{
(E^{\rm exact}_i - E^{\rm pred})^2/N}$, where $N$ is the number of
examples that we checked and here $N=200$)  is  $0.38$ MeV, assuming
that $d_0= D_0$, $k'=\frac{2}{5} k$ for all examples.

\begin{figure}
\includegraphics[angle=0,width=0.75\textwidth]{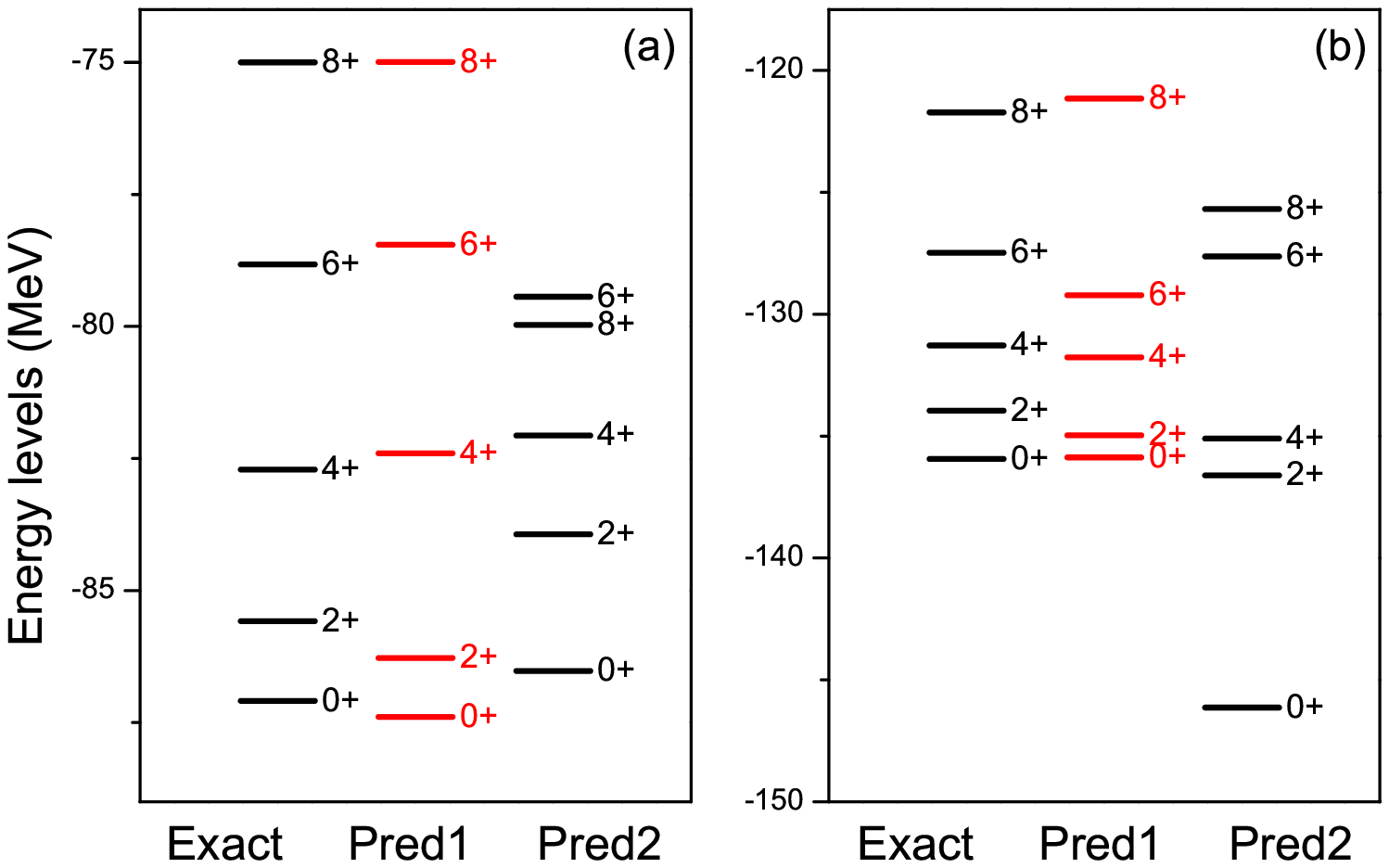}
\caption{~ (Color online) ~ Comparison of low-lying levels obtained
by exact diagonalization (denoted by ``exact")  and those predicted
by using Eq. (\ref{linear}) (denoted by ``pred1") and the linear
correlation formula suggested in Ref. \cite{Shen2} (denoted by
``pred2") for the $^{24}$Mg nucleus in panel (a), and the $^{28}$Si
nucleus in panel (b). The results are obtained by using the
Yukawa-type interaction of Refs. \cite{Arima,Takada4}. One sees
substantial improvement of predictions for the lowest eigenvalues
for each spin $I$ by using Eq. (1) of this paper.
 }\label{pred}
\end{figure}

Here we give a brief discussion of formulas in Refs.
\cite{V,Papenbrock1,Yoshinaga1,Shen1,Shen2} and the formula proposed
in this paper. Ref. \cite{Papenbrock1} reported  a correlation
between the lowest  eigenvalue and the spectral width $\sigma$ of
the spin $I$ states. Ref. \cite{V} suggested a simple formula
$E_{\rm min}^{(I)} = \bar{E}_I - \sqrt{ {\rm ln} D/{\rm ln} 2} ~
\sigma$, where $ \bar{E}_I$ is the average energy of spin $I$
states. Ref. \cite{Yoshinaga1} suggested a  formula $E_{\rm
min}^{(I)} = \bar{E}_I - \sqrt{ a{\rm ln} D+b}~ \sigma$ (similar to
Ref. \cite{V}), with $a\simeq 1.00$ and $b\simeq 0.40$. Ref.
\cite{Shen2} refined the results of Ref. \cite{Yoshinaga1} by
including the third moment analytically, with an additional factor
$\left( 1- \frac{\sqrt{\pi}}{6\sqrt{2}} \left(
\frac{\sigma_3}{\sigma} \right)^3 \right)$ multiplied in the second
term of the formula in Ref. \cite{Yoshinaga1}.  Here $\sigma_3$ is the
third moment of the eigenvalues. These formulas are applicable to
the random Hamiltonians statistically (i.e., the ensemble average), not
to the individual Hamiltonians such as for realistic systems.  In
Refs. \cite{Shen2}, the formula by using the linear correlation
between the eigenvalues and diagonal matrix elements is applicable
to individual sets of parameters and works well for medium
eigenvalues, but it does not work very well for the lowest (or the
largest) eigenvalues. The formula proposed in this paper is found to
predict remarkably well the lowest eigenvalues of the nuclear shell
model Hamiltonian, as shown in Fig. 4. We should also add that there
are many other efforts towards overcoming the limitation of
dimension in diagonalizing large matrices, see refs.
\cite{other-1,other-2,other-3,other-4}.

To summarize, in this paper we first investigated the distribution of
non-zero off-diagonal matrix elements of the nuclear shell model
Hamiltonian. We demonstrated that the non-zero off-diagonal matrix
elements exhibit regular patterns, if one sorts  the diagonal matrix
elements from the smaller to larger values; without sorting the
diagonal matrix elements, the off-diagonal matrix elements look
random. Almost all matrix elements becomes zero, if the matrix
elements are ``distant" enough from the diagonal line, after sorting
the diagonal matrix elements.

A very simple formula of the lowest eigenvalue for the shell model
Hamiltonian matrix $H^{(I)}$ is proposed, based on the regular
patterns of the $\rho(d)$ and $\epsilon_d$-ln$\,d$ plots for sub-matrices
$h$. There exists a ``critical" dimension, $D_0$, at which the slope
of the $\epsilon_d$-ln$\,d$ plot changes. The slope  for  $d>D_0$ is
empirically found to be equal to 2/5 of that for $d < D_0$ with
fluctuations. The value of $D_0$ is found to be equal to $d_0$ which
can be obtained easily by using $\rho(d_0)=0$. Here $\rho (d)$ represents
the probability for $H_{ij}$ to be non-zero while moving away from the
diagonal line.

The overall root-mean-squared deviation for two-hundred shell model
Hamiltonians of nuclei in the $sd$ shell is  $0.38$ MeV (the
relative deviation is about $0.003$), assuming that $d_0= D_0$ and
$k'=\frac{2}{5} k$, with $k$ and $b$ obtained from the sub-matrices
$h$ of $H^{(I)}$. The dimension of $h$ is much smaller than $D$, and
here we take $d\le D/10$. This demonstrates that our predicted
results of the lowest eigenvalue based on our new formula are in
very good agreement with the exact values, even if one treats much
smaller ``sub-matrices" of $H^{(I)}$. We therefore expect that our
new formula has significance for future theoretical studies of
nuclear structure. It will be also interesting to investigate
whether or not other low-lying states have similar features.

What has  not been yet understood at a microscopic level is why the
$\epsilon_d$-ln$\,d$  plot exhibits a remarkable linearity. Further
consideration of these issues is warranted in future studies.

{Acknowledgements:}   We thank the National Natural Science
Foundation of China for supporting this work under grant 10975096.
This work is also supported partly by Chinese Major State Basic
Research Developing Program under Grant 2007CB815000.

\end{document}